\documentclass[aps,twocolumn,pre,showpacs,superscriptaddress,amsmath,amssymb]{revtex4}  
\usepackage{graphicx,times,bm}

\newcommand{\nc}{\newcommand}
\nc{\comment}[1]{}

\newtheorem{theo}{Theorem}[section]
\newtheorem{defin}{Definition}[section]

\nc{\eps}{\varepsilon}
\nc{\dPdmu}{\frac{dP}{d\mu}}
\nc{\avg}[1]{\left\langle {#1} \right\rangle}
\nc{\shift}{(\Delta x_0,\Delta y_0)}

\nc{\zev}{\lambda_{0,j}^{(V)}} \nc{\fj}{f_{0,j}^{(V)}}

\nc{\both}{\lambda_0^{(1),(2)}} \nc{\evx}{\lambda_0^{(1)}}
\nc{\evy}{\lambda_0^{(2)}}

\nc{\lmin}{\lambda_{\rm min}} \nc{\fmin}{f_{\rm min}}
\nc{\lminV}{\lambda_{\rm min}^{(V)}} %
\nc{\fminV}{f_{\rm min}^{(V)}}

\nc{\fig}[4] { \begin{figure}[ht!]
    \centering{\scalebox{#1}{\includegraphics*{#2.eps}}}
    \caption{#4}
    \label{fig:#3}
  \end{figure}
}

\begin{document}

\title{Qualitative and quantitative analysis
of stability and instability dynamics
of positive lattice solitons}

\author{Y.~Sivan} \affiliation{Department of Physics and Astronomy,
Tel Aviv University, Tel Aviv 69978, Israel}

\author{G.~Fibich}
\affiliation{Department of Applied Mathematics, Tel Aviv University,
Tel Aviv 69978, Israel}

\author{B.~Ilan}
\affiliation{School of Natural Sciences, University of California,
Merced, P.O. Box 2039, Merced, California 95344, USA}

\author{M. I.~Weinstein}
\affiliation{Department of Applied Physics and Applied Mathematics, Columbia University,
New York, NY 10027 USA}

\begin{abstract}
We present a unified approach for qualitative and quantitative
analysis of stability and instability dynamics of positive bright
solitons in multi-dimensional focusing nonlinear media with a
potential (lattice), which can be periodic, periodic with defects,
quasiperiodic, single waveguide, etc. We show that when the soliton
is unstable, the type of instability dynamic that develops depends
on which of two stability conditions is violated. Specifically,
violation of the slope condition leads to a focusing instability,
whereas violation of the spectral condition leads to a
drift instability.
We also present a quantitative approach that allows to predict
the stability and instability strength.
\end{abstract}

 \maketitle

\section{Introduction}

Solitons, or solitary waves, are localized nonlinear waves that maintain
their shape during propagation. They are prevalent in many branches
of physics, and their properties have provided deep insight into
complex nonlinear systems.
The stability properties of solitons are of fundamental importance.
Stable solitons are both natural carriers of energy in naturally occurring systems and often the preferred carriers of energy in engineered systems.  Their stability  also makes them most  accessible to experimental observation.

The first studies considered stability of solitons in homogeneous
media. In recent years there has been a considerable interest in the
study of solitons in lattice-type systems. Such solitons have been
observed in optics using waveguide arrays, photo-refractive
materials, photonic crystal fibers, etc., in both one-dimensional
and multidimensional lattices, mostly periodic sinusoidal square
lattices~\cite{eisen-prl1998,efrem-pre2002pr,Christodoulides-03,Fleischer-03,Sukhorukov-03,Efremidis-03,Neshev-04,pertc-prl2004}
or single waveguide
potentials~\cite{carr-pra2001,Barad_single_wg,Tamura_sakagouchi_single_wg},
but also in discontinuous lattices (surface
solitons)~\cite{discrete_surface_solitons}, radially-symmetric
Bessel lattices~\cite{Kartashov-04}, lattices with triangular or
hexagonal symmetry~\cite{Kevrekidis-02,Kivshar-opt-lett-triangular},
lattices with
defects~\cite{Irregular-06,Fedele-05,Makasyuk-05,Martin-04,Qi-04,Ryu-03,Yang-06},
with quasicrystal
structures~\cite{Irregular-06,Bratfalean-05,Freedman-06,
Lifshitz-05,Man-05,Villa-05,Xie-03} or with random
potentials~\cite{Segev_andersson,Silberberg_andersson}. Solitons
have also been observed in the context of Bose-Einstein Condensates
(BEC)~\cite{Abdullaev_BEC_review,BEC_review}, where lattices have
been induced using a variety of techniques.

Stability of lattice solitons has been studied in hundreds of
papers. The majority of these papers focused on one specific
physical configuration, i.e., a specific dimension (mostly in 1D),
nonlinearity and
lattice type. In addition, in several studies, general conditions
for stability and instability were derived (see
Section~\ref{sec:history}). In all of these studies, the key
question was whether the soliton is stable (yes) or unstable (no).

Fibich, Sivan and Weinstein went beyond this binary view by
developing a {\em qualitative} and {\em quantitative} approach to
stability of positive lattice solitons. This was first carried out
for spatially non-homogeneous nonlinear potentials in
~\cite{NLS_NL_MS_1D,NLS_NL_MS_2D}. These ideas were then developed
by Sivan, Fibich and coworkers in the context of linear
non-homogeneous potentials in
\cite{narrow_lattice_solitons,delta_pot_complete,mobility}.
These studies showed that the
qualitative nature of the instability dynamics is determined by the
particular violated stability condition. In addition, they presented a
quantitative approach for  prediction of the stability or
instability strength. Specifically, these papers considered the
cases of a one-dimensional nonlinear lattice~\cite{NLS_NL_MS_1D}, a
two-dimensional nonlinear lattice~\cite{NLS_NL_MS_2D}, a
one-dimensional linear delta-function
potential~\cite{delta_pot_complete} and narrow solitons in a linear
lattice~\cite{narrow_lattice_solitons}.

In the present article, the results
of~\cite{NLS_NL_MS_1D,NLS_NL_MS_2D,narrow_lattice_solitons,
delta_pot_complete,mobility} are combined
into a unified theory for stability and instability of lattice
solitons that can be summarized in a few rules
(Section~\ref{sec:rules}).
We illustrate how these rules can be applied
in a variety of examples that may be useful to experimental studies.

\section{Model, notation and definitions}
We study the stability and instability dynamics
of lattice solitons of the nonlinear Schr\"odinger (NLS) equation
with an external potential, which in dimensionless form is given by
\begin{equation}
  \label{eq:NLS_l_nl_F}
  i A_z(\vec{x},z) + \Delta A + \left(1 - V_{nl}(\vec{x}) \right)F\left(|A|^2\right) A - V_l(\vec{x}) A = 0.
\end{equation}
Equation (\ref{eq:NLS_l_nl_F}) is also referred to as the Gross-Pitaevskii equation (GP).
NLS/GP underlies many models of nonlinear wave propagation in
nonlinear optics and macroscopic quantum systems (BEC). For example, in the context of laser
beam propagation, $A(\vec{x},z)$ corresponds to the electric field
amplitude, $z\ge0$ is the distance along the direction of propagation, $\vec{x} =
(x_1,\dots,x_d)$ is the transverse $d$-dimensional space [e.g., the
$(x,y)$ plane for propagating in bulk medium] and $\Delta =
\partial^2_{x_1} + \cdots + \partial^2_{x_d}$ is the $d$-dimensional diffraction term. The nonlinear term models the
intensity-dependence of the refractive index. For example,
$F\left(|A|\right) = |A|^2$ corresponds to the optical Kerr effect
and $F\left(|A|^2\right) = 1/\left(1 + |A|^2\right)$ corresponds to
photorefractive materials, see
e.g.,~\cite{photorefractive_solitons_review}. The potentials $V_l$
and $V_{nl}$ correspond to a modulation of the linear and nonlinear
refractive
indices, respectively. 
In BEC, $z=t$ is time, $A(\vec{x},t)$
represents the wave function of the mean-field atomic condensate,
$F\left(|A|^2\right) = |A|^2$ represents contact (cubic)
interaction, and the potentials $V_l(\vec x)$ and $V_{nl}(\vec x)$ are induced by
externally applied electro-magnetic fields~\cite{Pethick-01}.

We define a {\em soliton} to be any solution of Eq.~(\ref{eq:NLS_l_nl_F})
of the form $A(\vec{x},z) = u(\vec{x})e^{-i \mu z}$, where $\mu$ is
the propagation constant and $u(\vec{x})$, the soliton profile, is a
real-valued function that decays to zero at infinity
and satisfies
\begin{equation}
  \label{eq:bs}
  \Delta u + \left(1 - V_{nl}(\vec{x})\right)F(u^2) u + \mu u - V_l u = 0.
\end{equation}

Solitons can exist only  for $\mu$ in the gaps in the spectrum of the linear problem
\begin{equation}
  \label{eq:linear}
  \Delta u + \mu u - V_l u = 0,
\end{equation}
i.e., for values of $\mu$ such that the linear
problem~(\ref{eq:linear}) does not have any non-trivial solution,
see e.g.,~\cite{Ziad-Yang-04}.

Solitons in a lattice potential, or more general non-homogeneous
potential, may be understood as bounds states of an effective
(self-consistent) potential, $V_{eff}=V_l(\vec
x)+\left(-1+V_{nl}(\vec x)\right)F\left(u^2(\vec x)\right)$. They
arise (i) via bifurcation from the zero-amplitude state with energy
at an end point of a continuous spectral band (finite or
semi-infinite) of extended states of  the linear operator of
$-\Delta+V_l$ \cite{Kupper-Stuart:90}
or (ii) if
$V_l$ is a potential with a defect, via bifurcation from discrete
eigenvalues (localized linear modes) within the spectral gaps
(semi-infinite or finite), which in addition to the bare
nonlinearity, can serve to nucleate a localized nonlinear bound
state~\cite{Rose-Weinstein:88}.

In this paper, we only consider positive solitons ($u>0$) of both
type (i) and (ii). This is always the case for the least energy
state within the semi-infinite gap, i.e., when $-\infty < \mu <
\mu^{(V)}_{BE}$, where $\mu^{(V)}_{BE}$ is the lowest point in the
spectrum of Eq.~(\ref{eq:linear}), at which the first band begins.
Solitons whose frequencies lie in finite spectral gaps are usually
referred to as {\it gap solitons}. However, gap solitons typically
oscillate and change sign and are therefore not covered by the
theory presented in this paper~\footnote{Moreover, it can be shown
that for gap solitons $n_-(L_+)=\infty$ (see definition in
Section~\ref{sec:history}) hence they are not covered by
Theorem~\ref{th:stability}.}.

We study the dynamics of NLS/GP and its solitons in the space $H^1$, with norm $\|f\|^2_{H^1}:=\int(|f|^2+|\nabla f|^2|)d\vec{x}$. The natural notion of stability is orbital stability, defined as follows:
\begin{defin}
\label{defin:stability} Let $u(\vec{x})$ be a solution of
Eq.~(\ref{eq:bs}) with propagation constant $\mu$.  Then, the soliton solution $u(\vec{x})
e^{-i \mu z}$ of NLS/GP eqn. ~(\ref{eq:NLS_l_nl_F}) is
orbitally stable  if for all $\eps>0$, there exists $\delta(\eps) > 0$ such
that for any initial condition $A_0$ with
$\ \inf_{\gamma\in
\mathbb{R}}~\|A_0 - u e^{i\gamma} \|_{H^1} ~< ~\delta$, then for all $z\ge0$ the
corresponding solution $A(\vec x,z)$ of Eq.~(\ref{eq:NLS_l_nl_F}) satisfies
$ \inf_{\gamma\in \mathbb{R}} \|A(\cdot,z) - ue^{i\gamma} \|_{H^1} < \eps$ .
\end{defin}

In discussing the stability theory for NLS it is useful to refer to
its Hamiltonian structure:\ $i\partial_t A=\delta H / {\delta A^*}$,
where
\begin{equation}
H[A,A^*] = \int\left(|\nabla A|^2 + V_l|A|^2 - (1 -
V_{nl})G(|A|^2) \right) d\vec{x}, \nonumber
\end{equation}
and $G'(s)=F(s),\ G(0)=0$. The Hamiltonian, $H$, and the
optical power (particle number):
\begin{equation}
P  = \int |A|^2 d\vec{x} \nonumber
\end{equation}
are conserved integrals for NLS. Eq.~(\ref{eq:bs}) for $u(x;\mu)$,
the soliton profile, can be written equivalently as the energy
stationarity condition $\delta{\cal E}=0$, where $ {\cal E}\equiv H
- \mu P$.  Soliton stability requires a study of $\delta^2{\cal E}$,
the {\it second variational derivative} of ${\cal E}$ about $u$. For
NLS in general, stable solitons need to be {\it local} energy
minimizers; see {\it e.g.}~\cite{Rose-Weinstein:88} and
also~\footnote{Stability of solitons which are {\em global}
minimizers of the Hamiltonian, ${\cal H}$, subject to fixed squared
$L^2$ norm, $P$, was studied by Cazenave and
Lions~\cite{Cazenave-Lions-82}.   For $V\equiv0$ and power
nonlinearities, $F(|A|^2)=|A|^{2\sigma}$, a global minimizer (and
therefore stable soliton) exists  in  the subcritical case, $\sigma
< 2/d$. This condition on $\sigma$ is also a consequence of the
slope condition.}.

\section{Soliton stability -- overview}
\label{sec:history}

The first analytic result on soliton stability was obtained by
Vakhitov and Kolokolov~\cite{Vakhitov-73}. They proved, via a study
of the linearized perturbation equation, that a {\em necessary}
condition for {\em stability} of the soliton $u(x;\mu)$ is
\begin{equation}
\frac{d P(\mu)}{d\mu} < 0, 
 \label{slope}
\end{equation}
{\it i.e.} the soliton is stable only if its power decreases with
increasing propagation constant $\mu$. This condition will be
henceforth called the slope condition.

Subsequent studies of {\em nonlinear} stability analysis of solitons
revealed the central role played by the number of negative and
zero eigenvalues of the operator
\begin{equation}\label{eq:L+-}
L_+ = - \Delta + V_l - \left(1 - V_{nl}\right) \left(F(u^2) - 2 u^2
F'(u)\right)\ - \mu,
\end{equation}
which is the real part of $\delta^2{\cal E}$~\footnote{In order to
avoid confusion, we point out that the value of~$\mu$ in $L_+$ is
fixed, so that the eigenvalues and eigenfunctions of~$L_+$ are the
solutions of
$$
L_+(\mu;V) f(\vec{x}) = \lambda(\mu;V) f(\vec{x}).
$$}.
Weinstein~\cite{Weinstein-86} showed that for a homogeneous (translation-invariant) medium, ($V_l\equiv0$, $V_{nl}\equiv0$), for $u(\vec{x};\mu)>0$, if the
slope condition~(\ref{slope}) is satisfied and $L_+$ has only one negative eigenvalue, 
then the solitons are nonlinearly stable. Later, in~\cite{Rose-Weinstein:88} (Theorem 3.1; see also Theorem~6 of~\cite{Weinstein-89}) it was shown that in the presence of a linear potential which is bounded below and decaying
at infinity, solitons are stable if in addition, $L_+$ also has no zero eigenvalue(s). A related treatment was given
to the narrow soliton (semi-classical limit) subcritical nonlinearity case in~\cite{GSS,Oh-89,narrow_lattice_solitons} and for solitons in spatially varying nonlinear potentials in~\cite{NLS_NL_MS_1D,NLS_NL_MS_2D}.

General sufficient conditions for {\em instability} were given by
Grillakis~\cite{Grillakis:88} and Jones~\cite{Jones:88}. These
results imply that if either the slope is positive or if $L_+$ has
more than one negative eigenvalue, then the soliton is unstable.

A direct consequence of the arguments of Section~3 and Theorem~3.1
in~\cite{Rose-Weinstein:88},~\cite{Weinstein-89},
and~\cite{Jones:88,Grillakis:88}, is a stability theorem (used in
this paper), which applies to positive solitons of
NLS~(\ref{eq:NLS_l_nl_F}), whose frequencies lie in the
semi-infinite spectral gap of $-\Delta+V$; see
also~\cite{stability-theory-proof}.
\begin{theo}
\label{th:stability} Let $u(\vec{x})$ be a positive solution of
Eq.~(\ref{eq:bs}) with propagation constant $\mu$ within the
semi-infinite gap, {\it i.e.} $\mu<\mu_{BE}^{(V)}$. Then, $A =
u(\vec{x})e^{- i \mu z}$ is an orbitally-stable solution of the
NLS~(\ref{eq:NLS_l_nl_F}) if both of the following conditions hold:
\begin{enumerate}
\item
  The {\bf slope} (Vakhitov-Kolokolov) {\bf condition}:
  \begin{equation}
\label{eq:slope} \dPdmu < 0. 
\end{equation}
\item The {\bf spectral condition}: $L_+$ has no zero eigenvalues
and
  \begin{equation}
    \label{eq:spectral}
    n_-(L_+) = 1.
  \end{equation}
   \end{enumerate}
If either $\dPdmu > 0$ or $n_- \ge 2$, the soliton is unstable.
\end{theo}
We note that Theorem~\ref{th:stability} does not cover two cases:
\begin{enumerate}
\item $\dPdmu = 0$:\ For homogeneous media, $V_l = V_{nl} = 0$, solitons are unstable; see~\cite{Weinstein-83,Weinstein-85} for
power nonlinearities and~\cite{CoPe:03} for general nonlinearities.
There are no analytic results for inhomogeneous media.
\item {\it $n_-(L_+) = 1$ and zero is an eigenvalue of multiplicity one
or higher:}\ This case will be discussed in
Sections~\ref{sub:stability_hom_rev}
and~\ref{sub:stability_inhom_rev}.
\end{enumerate}

We also note that Grillakis, Shatah and Strauss
(GSS)~\cite{GSS,GSS2} gave an alternative abstract formulation of a
stability theory for positive solitons Hamiltonian systems,
including NLS with a general class of linear and nonlinear spatially
dependent potentials. In this formulation, the spectral condition on
$n_-(L_+)$ and the slope condition are {\it coupled}, see detailed
discussion in~\cite{narrow_lattice_solitons}. The formulation of
Theorem~\ref{th:stability} is a more refined and stronger statement.
Specifically, {\it it {\it decouples} the slope condition and the
spectral condition on $n_-(L_+)$ as two independent necessary
conditions for stability and shows that a violation of either of
them would lead to instability}. This decoupling is at the heart of
our qualitative approach since violation of each condition leads to
a different type of instability. Stability of solitons in
homogeneous media has also been investigated using the
Hamiltonian-Power curves, see e.g.,~\cite{referee-ref}.

\subsection{Review of stability conditions in homogeneous media}\label{sub:stability_hom_rev}

Stability and instability of solitons in \emph{homogeneous} media
(i.e., $V\equiv 0$) have been extensively
investigated~\cite{Sulem-99}. In this case, $\mu^{(V \equiv 0)}_{BE}
= 0$, i.e., the semi-infinite gap associated with
Eq.~(\ref{eq:linear}) is $(-\infty,0)$. For every $\mu<0$ and
$\vec{x}_0 \in \mathbb{R}^d$, there exists a soliton centered at
$\vec{x}_0$ which
is radially-symmetric in $r = |\vec{x}-\vec{x}_0|$, positive, and monotonically decaying in $r$.

In the case of a power-law nonlinearity $F(|u|) = |u|^{2\sigma}$, the slope condition~(\ref{eq:slope})
depends on the dimension $d$ and nonlinearity exponent~$\sigma$ as
follows~\cite{Weinstein-85,Weinstein-86}:
\begin{enumerate}
\item
  In the subcritical case $d<2/\sigma$, $\dPdmu<0$. Hence, the slope condition is satisfied.
\item
  In the critical case $d=2/\sigma$,
  the soliton power does not depend on $\mu$, i.e., $\dPdmu \equiv 0$.
  By~\cite{Weinstein-86}, the slope condition is violated.
\item
  In the supercritical case $d>2/\sigma$, $\dPdmu>0$.
  Hence, the slope condition is violated.
\end{enumerate}
Thus, the slope condition is satisfied only in the subcritical case.

When $V\equiv 0$, the spectrum of $L_+$ is comprised of three
(essential) parts~\cite{Weinstein-85}, see Figure~\ref{fig:spectrum1}:\\
\begin{enumerate}
\item
  A simple negative eigenvalue $\lmin<0$,
  with a corresponding positive and radially-symmetric
  eigenfunction~$\fmin$. In~\cite{narrow_lattice_solitons}, it was shown that for power nonlinearities, $F(|u|) = |u|^{2\sigma}$, $\fmin~=~u^{\sigma+1}$ and $\lambda_{min} = \sigma(\sigma+2)\mu$.
\item
  A zero eigenvalue with multiplicity $d$,
  i.e., $\lambda_{0,j} = 0$ with eigenfunctions $f_j = \frac{\partial u}{\partial x_j}$ for $j=1,\dots,d$.
  These zero eigenvalues manifest the translation invariance in
  a homogeneous medium in all $d$ directions.
\item
  A strictly positive continuous spectrum $[-\mu,\infty)$.
\end{enumerate}

\fig{0.4}{spectrum1}{spectrum1}{
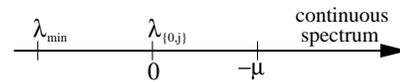
   The spectrum of $L_+$ in a homogeneous medium. }

Theorem~\ref{th:stability} does not apply directly for the stability
of solitons in homogeneous medium because $\lambda_{0,j} = 0$ and
$n_- = 1$. Accordingly, the notion of orbital stability must be
modified. Indeed, by the Galilean invariance of NLS for
$V_l=V_{nl}=0$, an arbitrarily small perturbation of a soliton can
result in the soliton moving at small uniform speed to infinity. The
orbit  in a homogeneous medium is thus the group of all translates
in phase and space, \emph{i.e.,} $\{u(\vec{x}-\vec{x}_0;\mu)
e^{i\gamma}:\ \vec{x}_0\in\mathbb{R}^d,\ \gamma\in[0,2\pi)\ \}$ and
orbital stability is given by Definition~\ref{defin:stability} but
where the infimums are taken over all $\gamma$ and $\vec{x}_0$.

Accordingly, Weinstein showed in~\cite{Weinstein-86} that in the
case of homogeneous media, the spectral condition can be slightly
relaxed so that it is satisfied if $L_+$ has only one {\em negative}
eigenvalue and $d-$ zero eigenvalues, associated with the
translational degrees of freedom of NLS. Hence, the spectral
condition is satisfied in homogeneous media and stability is
determined by the slope condition alone~\cite{Weinstein-86}. In
particular, solitons in homogeneous media with a power-law
nonlinearity $F(|A|^2)=|A|^{2\sigma}$ are stable only in the
subcritical case $\sigma<2/d$.

\subsection{Stability conditions in inhomogeneous
media}\label{sub:stability_inhom_rev}

Below we investigate how the two stability conditions are affected by a potential/lattice.

Generically, in the subcritical ($d < 2/\sigma$) and supercritical
($d > 2/\sigma$) cases, the slope has an $\mathcal{O}(1)$ magnitude
in a homogeneous medium. Hence, a weak lattice can affect the
magnitude of the slope but not its sign, see
e.g.,~\cite{narrow_lattice_solitons}. Clearly, a sufficiently strong
lattice can alter the sign of the slope, see
e.g.~\cite{delta_pot_complete} for the subcritical case and
~\cite{mihal-pra2005,mihal-3d-optics} for the supercritical case.
The situation is very different in the critical case ($d =
2/\sigma$). Indeed, since the slope is zero in a homogeneous medium,
any potential, no matter how weak, can affect the sign
of the slope.

The potential can affect the spectrum of $L_+$ in two different
ways: 1) shift the eigenvalues, and 2) open gaps (bounded-intervals)
in the continuous spectrum, see Figure~\ref{fig:spectrum2}. In
general, the minimal eigenvalue of $L_+$ remains negative, i.e.,
$\lminV<0$, the
continuous spectrum remains positive, and the zero eigenvalues can move either to the right or to the left.
Hence, generically, the spectrum of $L_+$ has the following
structure:
\begin{enumerate}
\item A simple negative eigenvalue $\lminV<0$ with a positive
eigenfunction $\fminV>0$.

\item
  Perturbed-zero eigenvalues $\zev$ with eigenfunctions $f_j^{(V)}$, for $j = 1, \dots, d$.

\item
  A positive continuous spectrum, sometimes with a band-gap structure,
  beginning at $-\mu^{(V)}_{BE} > 0$.
\end{enumerate}
This structure of the spectrum was proved in~\cite{NLS_NL_MS_1D} for
solitons in the presence of a \emph{nonlinear lattice}, i.e.,
Eq.~(\ref{eq:NLS_l_nl_F}) with $V_l \equiv 0$. For a \emph{linear lattice}, the
proof of the negativity of $\lminV$ is the same as
in~\cite{NLS_NL_MS_1D}. The proof of the positivity of
$-\mu^{(V)}_{BE}$ is the same as in~\cite{NLS_NL_MS_1D} for
potentials that decay to $0$ as $|\vec{x}| \to \infty$.

\fig{0.4}{spectrum2}{spectrum2}{
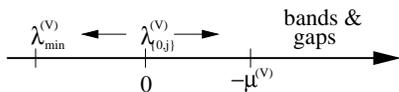
   The spectrum of $L_+$ in an inhomogeneous medium.
  }

Since $\lminV<0$ and the continuous spectrum is positive, the
spectral condition~(\ref{eq:spectral}) reduces to
\begin{equation}
  \label{eq:spectral2}
  \zev > 0~, \quad \quad j=1, \dots, d,
\end{equation}
i.e., that all the perturbed-zero eigenvalues are positive.
Generically, the equivalent spectral condition~(\ref{eq:spectral2})
is satisfied when the soliton is centered at a local minimum of the
potential, but violated when the soliton is centered at a local
maximum or saddle point of the
potential~\cite{Oh-89,Pelinovsky-04,NLS_NL_MS_1D,NLS_NL_MS_2D,
mobility,delta_pot_complete,narrow_lattice_solitons,Rapti_Kev_jones,Lin_Wei}.

Although generically $\zev(\mu) \ne 0$, there are two scenarios in
which $\zev$ equals zero:
\begin{enumerate}
  \item The potential is invariant under a {\em subgroup} of the continuous
spatial-translation group. For example (see
also~\cite{NLS_NL_MS_2D}), in a one-dimensional lattice embedded in
2D, i.e., $\frac{\partial V(x,y)}{\partial y}\equiv 0$, one has
$\lambda_{0,2}(\mu)\equiv 0$. In such cases, the zero eigenvalues do
not lead to instability for the reasons given in
Section~\ref{sub:stability_hom_rev}. Rather, the orbit and distance
function are redefined modulo the additional invariance, e.g., in
the example above the orbit is $\{u(x,y-y_0;\mu) e^{i\gamma}:\
y_0\in\mathbb{R},\ \gamma\in[0,2\pi)\ \}$. 
\item In the presence of spatial inhomogeneity, $V\ne0$, $\zev$ can cross zero as $\mu$ is varied. See for example~\cite{KKSW:08} and
the examples discussed in Sections~\ref{sub:shallow_max}
and~\ref{sec:defects}. This crossing can be associated with a
bifurcation and the existence of a new branch of solitons and an
exchange of stability from the old to the new branch; see the symmetry breaking analysis of ~\cite{KKSW:08}.
In such cases, stability and instability depend  on the details of
the potential and nonlinearity.
\end{enumerate}

In some cases, there are also positive discrete eigenvalues in
$(0,-\mu)$. 
However, these eigenvalues do not affect the orbital
stability, since they are positive. They do play a role, however, in
the scattering theory of solitons~\cite{SW-RMP:04,SW-PRL:05}.

We note that in many previous studies, only the slope condition was
checked for stability. As Theorem~\ref{th:stability} shows, however,
``ignoring'' the spectral condition is justified only for solitons
centered at lattice minima, since only then the spectral condition
is satisfied. In all other cases, checking only the slope condition
usually lead to incorrect conclusions regarding stability.

\subsection{Instability and collapse}\label{sub:instability_collapse}
We recall that in a homogeneous medium with a power nonlinearity,
all solutions of the subcritical NLS exist globally. For critical
and supercritical NLS there are collapsing (singular)
solutions~\cite{Weinstein-83}, i.e., solutions for which $\int
|\nabla A(\vec x,z)|^2\ d\vec x$ tends to infinity in finite
distance. Hence, in a homogeneous medium, the two phenomena of
collapse and of soliton instability appear together. In fact, the
two phenomena are directly related, since in the critical and
supercritical cases, the instability of the solitons is manifested
by the fact that they can collapse under infinitesimally small
perturbations (i.e., a {\em strong instability}).

As we shall see below, the situation is different in inhomogeneous
media. Indeed, the soliton can be unstable even if all solutions of
the corresponding NLS exist globally. Conversely, the soliton can be
stable, yet undergo collapse under a sufficiently strong
perturbation. Such results on the ``decoupling'' of instability and
collapse have already appeared
in~\cite{NLS_NL_MS_1D,NLS_NL_MS_2D,Irregular-06,narrow_lattice_solitons,delta_pot_complete,bounded-01}.
In all of these cases, the ``decoupling'' is related to the absence
of translation invariance.

\section{Qualitative approach -- Classification of instability
dynamics} \label{sec:qualitative}

The dynamics of orbitally-stable solitons is relatively
straightforward~- the solution remains close to the unperturbed
soliton. On the other hand, there are several possible ways for a
soliton to become unstable: it can undergo collapse, complete
diffraction, drift, breakup into separate structures, etc.

Theorem~\ref{th:stability} is our starting point for the
classification of the instability dynamics,
since it suggests that there are two independent
mechanisms for (in)stability. In fact, we show below that
the instability dynamics depends on which of the two
conditions for stability is violated.

As noted in Section~\ref{sub:instability_collapse}, in a homogeneous
medium with a power-law nonlinearity, when the slope condition is
violated, the soliton can collapse (become singular) under an
infinitesimal perturbation. If the perturbation increases the beam
power, then nonlinearity dominates over diffraction so that the
soliton amplitude becomes infinite as its width shrinks to zero. If
the perturbation is in the ``opposite direction'', the soliton
diffracts to zero, i.e., its amplitude goes to zero as its width
becomes infinite, see e.g., Theorem 2 of~\cite{Weinstein-89}. More
generally, in other types of nonlinearities or in the presence of
inhomogeneities, there are cases where the slope condition is
violated but collapse is not possible (e.g., in the one-dimensional
NLS with a saturable nonlinearity~\cite{Malomed-CQ1D}). In such
cases, a violation of the slope condition leads to a focusing
instability whereby infinitesimal changes of the soliton can result
in large changes of the beam amplitude/width, but not in collapse or
total diffraction. Accordingly, we refer to the instability which is
related to the violation of the slope condition as a {\em focusing
instability} (rather than as a collapse instability).

When the soliton is unstable because the spectral condition is
violated, it undergoes a {\em drift instability} whereby
infinitesimal shifts of the initial soliton location lead to a
lateral movement of the soliton away from its initial location. The
mathematical explanation for the drift instability is as follows.
The spectral condition is associated with the perturbed-zero
eigenvalue $\lambda_{0,j}^{(V)}$ and the corresponding eigenmode
$f_j$. In the homogeneous case, the eigenmodes $f_j = \frac{\partial
u}{\partial x_j}$ are odd. By continuity from the homogeneous case,
the perturbed-zero eigenmodes $f_j^{(V)}$ in the presence of a
potential are odd for symmetric potentials and ``essentially'' odd
for {\em asymmetric} potentials. When the spectral condition is
violated, these odd eigenmodes grow as $z$ increases, resulting in an
\emph{asymmetric} distortion of the soliton, which gives rise to a
drift of the beam away from its initial location. The mathematical relation between
the violation of the spectral condition and the drift instability is
further developed in Section~\ref{sec:quantitative}.

Finally, the drift dynamics also has an intuitive physical explanation.
According to Fermat's Principle, light bends towards regions of
higher refractive-index. Positive values of the potential~$V$
correspond to negative values of the refractive index, hence, Fermat's
principle implies that beams bend towards regions of lower
potential. Moreover, since generically, the spectral condition is satisfied for solitons
centered at a lattice minimum but violated for solitons centered at
a lattice maximum, one sees that the drift instability of solitons
centered at lattice maxima and the drift stability of solitons
centered at lattice minima is a manifestation of Fermat's principle.

\section{Quantitative approach}\label{sec:quantitative}
As noted, the soliton is drift-unstable when $\zev < 0$ but
drift-stable when $\zev \ge 0$. Thus, there is a discontinuity in
the behavior as $\zev$ passes through zero. Nevertheless, one can
expect the transition between drift instability and drift stability
to be continuous, in the sense that as $\zev$ approaches zero from
below, the rate of the drift becomes slower and slower. Similarly,
we can expect that as $\zev$ becomes more negative, the drift rate will increase.

The quantitative relation between the value of $\zev$ and the drift
rate was found \emph{analytically} for the first time
in~\cite{narrow_lattice_solitons} for {\em narrow} solitons in a
Kerr medium with a linear lattice. Later, based on the linearized
NLS dynamics, it was shown in~\cite{mobility} that for solitons of
any width, any nonlinearity and any linear or nonlinear potential,
this quantitative relation is as follows. Let us define the center
of mass of a perturbed soliton in the $x_j$~coordinate as
\begin{equation}
  \label{eq:COM}
 \avg{x_j} := \frac{1}{P} \int x_j |A|^2 d\vec{x}.
\end{equation}
Then, by~\cite{mobility}, the dynamics of $\avg{x_j}$ is initially governed by the
linear oscillator equation
\begin{equation}
  \label{eq:drift}
  \frac{d^2}{dz^2}\left(\avg{x_j} - \xi_{0,j}\right) = \Omega_j^2 \left(\avg{x_j} - \xi_{0,j}\right),
  \end{equation}
with the initial conditions
\begin{equation}
\left\{
\begin{array}{llll}
\avg{x_j}_{z=0} = \int x_j |A_0|^2 d\vec{x} / P, \\ \\
\frac{d}{dz}\avg{x_j}_{z=0} = 2 d \cdot Im \int A_0^* \nabla A_0 d\vec{x}
/ P.
\end{array}
\right. \label{eq:drift_IC}
\end{equation}
Here, $\xi_{0,j}$ is the location of the lattice critical point in the $j$th
direction (not to be confused with $\avg{x_j}_{z=0}$, the value of
the center of mass at $z=0$). The forcing is given by
\begin{equation} \label{eq:Omega}
\Omega_j^2 = - C_j\zev, \quad C_j =
\dfrac{(\fj,\fj)}{(L_-^{-1}\fj,\fj)},
\end{equation}
where $\fj$ is the eigenmode of $L_+$ that corresponds to $\zev$, i.e.,
the eigenmode along the $x_j$ direction, the operator $L_-$ is given
by
$$
L_- = - \Delta - \mu - \left(1 - V_{nl}(\vec{x})\right) F(u^2) +
V_l,
$$
and the inner product is defined as $(f,g)=\int fg^*d\vec{x}$.

Since $L_-$ is non-negative for positive solitons, it follows that
$C_j>0$. Therefore, when $\zev$ is negative, $\Omega_j$ is real and
when $\zev$ is positive, $\Omega_j$ is purely imaginary. Hence, by
Eqs.~(\ref{eq:drift})-(\ref{eq:Omega}), it follows that the lateral
dynamics of a general incident beam centered near a lattice minimum
is
\begin{equation}
\avg{x_j} = \avg{x_j}_{z=0} \cos(|\Omega_j| z) +
\frac{\frac{d}{dz}\avg{x_j}_{z=0}}{|\Omega_j|} \sin(|\Omega_j| z),
\end{equation}
i.e., the soliton drifts along the $x_j$ coordinate at the rate
$\Omega_j$. On the other hand, the lateral dynamics of a general
incident beam centered near a lattice maximum is
\begin{equation}
\avg{x_j} = \avg{x_j}_{z=0} \cosh(\Omega_j z) +
\frac{\frac{d}{dz}\avg{x_j}_{z=0}}{\Omega_j} \sinh(\Omega_j z).
\label{eq:com_dyn_sol}
\end{equation}
i.e., the soliton is pulled back towards $\xi_{0,j}$ by a restoring
force which is proportional to $\Omega_j^2$, so that it undergoes
oscillations around $\xi_{0,j}$ in the $x_j$ coordinate with the
period $|\Omega_j|$.

As noted, the soliton is focusing-unstable when the slope $dP/d\mu$
is non-negative, and focusing-stable when the slope is negative. In
a similar manner to the continuous transition between drift
stability and instability, one can expect the transition between
focusing stability and instability to be continuous. In other
words, one can expect the magnitude of the slope to be related to
the strength of focusing stability or instability. At present, the
quantitative relation between the magnitude of the slope and the
strength of the stability is not known, i.e., we do not have a
relation such as~(\ref{eq:drift}). However, numerical evidence for
this link was found in several of our earlier
studies~\cite{NLS_NL_MS_1D,NLS_NL_MS_2D,narrow_lattice_solitons,delta_pot_complete}.
For example, in the case of focusing-stable solitons that collapse
under sufficiently large perturbations, it was observed that as the
magnitude of the slope increases, the magnitude of the perturbation
that is needed for the soliton to collapse also increases. Thus, the
magnitude of the slope is related to the size of the basin of
stability~\cite{NLS_NL_MS_1D,NLS_NL_MS_2D,narrow_lattice_solitons}.
In cases of focusing-stable solitons where collapse is not
possible, when the magnitude of the slope increases, the focusing
stability is stronger in the sense that for a given perturbation,
the maximal deviation of the soliton from its initial amplitude
decreases~\cite{delta_pot_complete}.

\subsection{Physical vs. Mathematical stability} The quantitative
approach is especially important in the limiting cases of ``weak
stability/instability'', {\it i.e.} when one is near the transition
between stability and instability.  For example, consider a soliton
for which the two conditions for stability are met, but for which
$\zev$ or the slope are very small in magnitude. Such a soliton is
orbitally stable, yet it can become unstable under perturbations
which are quite small compared with typical perturbations that exist
in experimental setups. Hence, such a soliton is ``mathematically
stable'' but ``physically unstable'', see e.g.,~\cite{NLS_NL_MS_1D}.
Conversely, consider an unstable soliton for which either $\zev$ is
negative but very small in magnitude or the slope is positive but
small. In this case, the instability develops so slowly so that it
can be sometimes neglected over the propagation distances of the
experiment. Such a soliton is therefore ``mathematically unstable''
but ``physically stable''~\cite{narrow_lattice_solitons}.

\section{General rules}\label{sec:rules}
We can summarize the results described so far by several general
rules for stability and instability of bright positive lattice
solitons.

The {\em qualitative approach} rules are:
\begin{enumerate}
\item[QL1] Bright positive lattice solitons of NLS equations can become
unstable in {\em only} two ways: focusing-instability or
drift-instability.
\item[QL2]
  Violation of the slope condition leads to an focusing-instability, i.e., either initial diffraction
  or initial self-focusing. In the latter case, self-focusing can lead to collapse. Note, however, that for ``subcritical'' nonlinearities, the self-focusing is
  arrested.
\item[QL3]
  The spectral condition is generically satisfied when the soliton is centered at a potential minimum
  and violated when the soliton is centered at a potential maximum or saddle point.
\item[QL4]
  Violation of the spectral condition leads to a drift-instability, i.e., an initial lateral drift of the
  soliton from the potential maximum/saddle point towards a nearby lattice minimum.
\end{enumerate}
The {\em quantitative theory} rules are:
\begin{enumerate}
\item[QN1]
  The strength of the focusing- and drift- stability and instability depends on the magnitude of
  the slope $\left|\dPdmu\right|$ and the magnitude of $|\zev|$, respectively.
\item[QN2]
  The lateral dynamics of the beam is initially given by
  Eqs.~(\ref{eq:drift})-(\ref{eq:Omega}).
\end{enumerate}
The above rules were previously demonstrated for 1D solitons in a
periodic nonlinear lattice~\cite{NLS_NL_MS_1D}, for an anisotropic
2D lattice~\cite{NLS_NL_MS_2D} and for several specific cases of
linear lattices~\cite{narrow_lattice_solitons,delta_pot_complete}.
In this paper, we demonstrate that these rules apply in a {\em
general setting} of dimension, nonlinearity, linear/nonlinear
lattice with any structure and for any soliton width. In particular,
we use these general rules to explain the dynamics of lattice
solitons in a variety of examples that were not studied before.

\section{Numerical methodology}\label{sec:methodology}

Below we present a series of numerical computations that illustrate
the qualitative and quantitative approaches presented in
Sections~\ref{sec:qualitative}-\ref{sec:quantitative}. We present
results for the 2D cubic NLS
\begin{equation}
\label{eq:NLS}
   i A_z(x,y,z) + \Delta A + |A|^2 A - V(x,y)A = 0,
\end{equation}
with periodic lattices, lattices with a vacancy defect, and lattices
with a quasicrystal structure. There are two reasons for the choice
of the 2D cubic NLS. First, this equation enables us to illustrate
the instability dynamics in dimensions larger than one, in
particular, in cases where the dynamics in each direction is
different (e.g., as for solitons centered at saddle points). Second,
the 2D cubic NLS enables us to elucidate the distinction between
instability and collapse. Indeed, we recall that a necessary
condition for collapse in the 2D cubic NLS is that the power of the
beam exceeds the critical power $P_c \approx 11.7$~\cite{Weinstein-83}.

We first compute the soliton profile by solving Eq.~(\ref{eq:bs})
using the spectral renormalization method~\cite{MJA-Ziad-05}. Once
the solitons are computed for a range of values of $\mu$, the slope
condition~(\ref{eq:slope}) is straightforward to check. In order to
check the spectral condition~(\ref{eq:spectral}), the perturbed-zero
eigenvalues $\zev$ (and the corresponding eigenfunctions $f_j$) of
the discrete approximation of the operator $L_+$ are computed using
the numerical method presented
in~\cite[Appendix~D]{narrow_lattice_solitons}. The value of
$\Omega_j$ is calculated from Eq.~(\ref{eq:Omega}) by inversion of
the discrete approximation of the operator $L_-$.

Eq.~(\ref{eq:NLS}) is solved using an explicit Runge-Kutta
four-order finite-difference scheme.
Following~\cite{NLS_NL_MS_1D,NLS_NL_MS_2D,narrow_lattice_solitons,delta_pot_complete},
the initial conditions are taken to be the unperturbed lattice
soliton $u(x,y)$ with \emph{either}
\begin{enumerate}
  \item a small {\em power perturbation},
  i.e.,
\begin{equation}
  \label{eq:sym_pert}
  A_0(x,y) = \sqrt{1+c}\, u(x,y),
\end{equation}
where $c$ is a small constant that expresses the excess power of the
input beam above that of the unperturbed soliton, \emph{or}
\item a small {\em lateral shift}, i.e.,
\begin{equation}
  \label{eq:asym_pert}
  A_0(x,y)=u(x-\Delta x_0,y-\Delta y_0),
\end{equation}
where $\Delta x_0$ and $\Delta y_0$ are small compared with the
characteristic length-scale (e.g., period) of the potential.

\end{enumerate}
The motivation for this choice of perturbations is that each
perturbation {\em predominantly} excites only one type of
instability. Indeed, by Eq.~(\ref{eq:drift})-(\ref{eq:drift_IC}), it
is easy to verify that under a power
perturbation~(\ref{eq:sym_pert}), the center of mass will remain at
its initial location
(cf.~\cite{NLS_NL_MS_1D,NLS_NL_MS_2D,delta_pot_complete}), i.e., no
lateral drift will occur. In this case, only an focusing instability
is possible. On the other hand, the asymmetric
perturbation~(\ref{eq:asym_pert}) will predominantly excite a drift
instability (but if the soliton is drift-stable, this perturbation
can excite an focusing instability, see
Figure~\ref{fig:DNS_cosmin_shift}).

The advantage of the
perturbations~(\ref{eq:sym_pert})-(\ref{eq:asym_pert}) over adding
random noise to the input soliton is that they allow us to control
the type of instability that is excited. Moreover, grid convergence
tests are also simpler. Once the NLS solution is computed, it is
checked for focusing and drift instabilities by monitoring the
evolution of the normalized peak intensity
\begin{equation}
  \label{eq:peak}
  I(z) := \frac{max_{x,y}|A(x,y,z)|^2}{|A_0(x,y)|^2},
\end{equation}
and of the center of mass~(\ref{eq:COM}), respectively.

\section{Periodic square lattices}
\label{sec:periodic}

We first choose the sinusoidal square lattice
\begin{equation}
  \label{eq:coscos}
  V(x,y) = \frac{V_0}{2}\left[\cos^2(2\pi x) + \cos^2(2\pi y) \right],
\end{equation}
which is depicted in Figure~\ref{fig:coscos}. We consider this to be
the simplest 2D periodic potential, as all the local extrema are
also global extrema. This lattice can be created through
interference of two pairs of counter-propagating plane waves, and is
standard in experimental setups, see,
e.g.,~\cite{cos_squared_exp1,cos_squared_exp2}. The stability and
instability dynamics are investigated below for solitons centered at
the lattice maxima, minima, and saddle points, see
Figure~\ref{fig:coscos}(b).

\fig{0.45}{coscos}{coscos}{
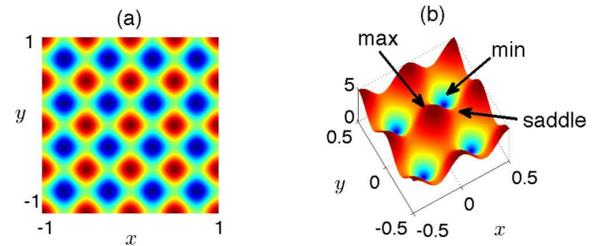
  (Color online) The sinusoidal square lattice given by Eq.~(\ref{eq:coscos})
  with $V_0=5$. (a) Top view. (b) Side view.
  The solitons investigated below are centered at
  the lattice maximum (0,0), lattice minimum (0.25,0.25),
  and saddle point (0.25,0).}

\subsection{Solitons at lattice minima}
\label{sub:coscos_min}

We first investigate solitons centered at the lattice minimum
$(x_0,y_0)=(0.25,0.25)$. Figure~\ref{fig:cos_stability_mu}(a) shows
that the power of solitons at lattice minima is below the critical
power for collapse, i.e., $P(\mu)<P_c\approx 11.7$ for all $\mu$. As
the soliton becomes narrower ($\mu \to -\infty$), the soliton power
approaches~$P_c$ from below (as was shown numerically
in~\cite{Ziad-Yang-04} for this lattice and analytically
in~\cite{narrow_lattice_solitons} for {\em any} linear lattice). In
addition, as the soliton becomes wider ($\mu \to \mu^{(V)}_{BE}$,
the edge of the first band), its power approaches $P_c$ from below
(rather than becomes infinite, as implied in~\cite{Ziad-Yang-04}), see also~\footnote{In fact, the soliton power approaches $g P_c$ where $g<1$, see~\cite{wide_solitons}. }.
The minimal power is obtained at $\mu = \mu_m \cong -10$. The power
curve thus has a stable branch for narrow solitons ($-\infty < \mu <
\mu_m$) where the slope condition is satisfied, and an unstable
branch for wide solitons ($\mu_m < \mu < \mu^{(V)}_{BE}$) where the
slope condition is violated. Therefore, wide solitons should be
focusing-unstable while narrow solitons should be focusing-stable.
Figure~\ref{fig:cos_stability_mu}(b) shows that, as expected for
solitons at lattice minima, $\lambda_0^{(1)} = \lambda_0^{(2)} > 0$
for all $\mu$. Hence, the spectral condition is fulfilled.
Consequently, solitons at lattice minima should not experience a
drift instability.

In order to excite the focusing instability alone, we add to the
soliton a small power perturbation, see Eq.~(\ref{eq:sym_pert}). We
contrast the dynamics in a neighborhood of stable and unstable
solitons by choosing two solitons with the same power ($P \cong 0.98
P_c$), from the stable branch ($\mu = -31$) and from the unstable
branch ($\mu = -3$). We perturb these solitons with the same power
perturbations ($c = 0.5\%, 1\%, 2\%$).

When $c = 0.5\%$ and $1\%$, the input power is below the threshold
for collapse ($P<P_c$). In these cases, the self-focusing process is
arrested and, during further propagation, the normalized peak
intensity undergoes oscillations (see
Figures~\ref{fig:DNS_cosmin}(a) and (b)). For a given perturbation,
the oscillations are significantly smaller for the stable soliton
compared with the unstable soliton.

When $c = 2.5\%$, the input power is above the threshold for collapse
($P>P_c$) and the solutions undergo collapse. Therefore, for such
large perturbations, collapse occurs for both stable and unstable
solitons, i.e., even when \emph{both the slope and spectral
conditions are fulfilled}.
This shows yet again that in an inhomogeneous medium, collapse and
instability
are not necessarily correlated.

\fig{0.5}{cos_stability_mu}{cos_stability_mu}{
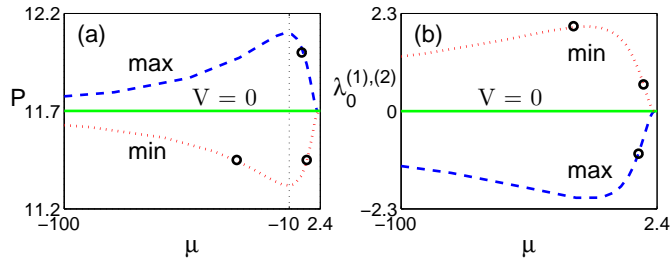
  (Color online) (a) Power, and (b) perturbed-zero eigenvalues, as functions of
    the propagation constant, for solitons centered at a
    maximum (blue, dashes) and minimum (red, dots) of the
    lattice~(\ref{eq:coscos}) with $V_0=5$.
    Also shown are the corresponding lines for the
    homogeneous NLS equation (solid, green).
    The circles (black) correspond to the values used
    in Figs.~\ref{fig:DNS_cosmin}-\ref{fig:DNS_cosmax}.
  }

\fig{0.45}{DNS_cosmin}{DNS_cosmin}{
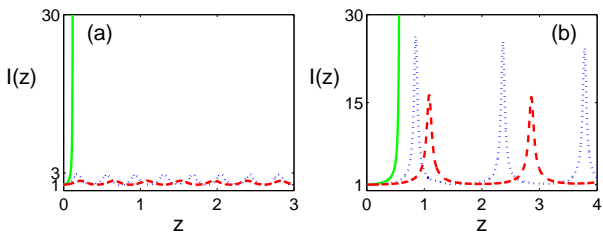
    (Color online) Normalized peak intensity~(\ref{eq:peak})
    of solutions of Eq.~(\ref{eq:NLS})
    with the
    periodic lattice~(\ref{eq:coscos}) with $V_0=5$. Initial conditions are
    power-perturbed solitons [see Eq.~(\ref{eq:sym_pert})] centered at a lattice minimum:
    (a) Soliton from the stable branch ($\mu = -31$);
    (b) Soliton from the unstable branch ($\mu = -3$).
    Input powers are $0.5\%$ (red dots), $1\%$ (blue dashes),
    and $2.5\%$ (solid green) above the soliton power.
  }

In order to confirm that solitons centered at a lattice minimum do
not undergo a drift instability, we shift the soliton slightly
upward by using the initial condition~(\ref{eq:asym_pert}) with
$\shift=(0,0.04)$. Under this perturbation, the solution of
Eq.~(\ref{eq:drift}) is
\begin{equation}\label{eq:coscos_min}
\avg{x} \equiv 0, \quad \quad \avg{y} = \Delta y_0 \cdot \cos
(|\Omega_y| z).
\end{equation}
In addition, by Eq.~(\ref{eq:Omega}), $\Omega_y \approx 11.12i$ for
$\mu = -31$ and $\Omega_y \approx 2.58i$ for $\mu = -3$.
Figure~\ref{fig:DNS_cosmin_shift}(a1) shows that for $\mu = -31$,
the center of mass in the $y$-direction of the position-shifted
soliton follows the theoretical prediction~(\ref{eq:coscos_min})
accurately over several oscillations. In addition, the center of
mass in the $x$-direction remain at $x=0$ (data not shown), in
agreement with Eq.~(\ref{eq:coscos_min}). Thus, the soliton is
indeed drift-stable.

The situation is more complex for $\mu = -3$. In this case, the
position-shifted soliton follows the theoretical
prediction~(\ref{eq:coscos_min}) over more than $2$ diffraction
lengths (i.e., for $z > z_0$ where $z_0 \approx 1$), but then
deviates from it, see Figure~\ref{fig:DNS_cosmin_shift}(b1). The
reason for this instability is that for $\mu = -3$, the slope
condition is violated. Since the position-shifted initial condition
can also be viewed as an asymmetric amplitude power perturbation
$\Delta A = u(x - \Delta x_0,y - \Delta y_0) - u(x,y)$, an focusing
instability is excited and the soliton amplitude decreases (as its
width increases), see Figure~\ref{fig:DNS_cosmin_shift}(b2).
Obviously, once the soliton amplitude changes significantly, the
theoretical prediction for the lateral dynamics is no longer valid.
In order to be convinced that the initial instability in this case
is of an focusing-type rather than drift-type, we note that for $\mu
= -31$ for which the slope condition is satisfied, the soliton
remains focusing-stable, see Figure~\ref{fig:DNS_cosmin_shift}(a2).

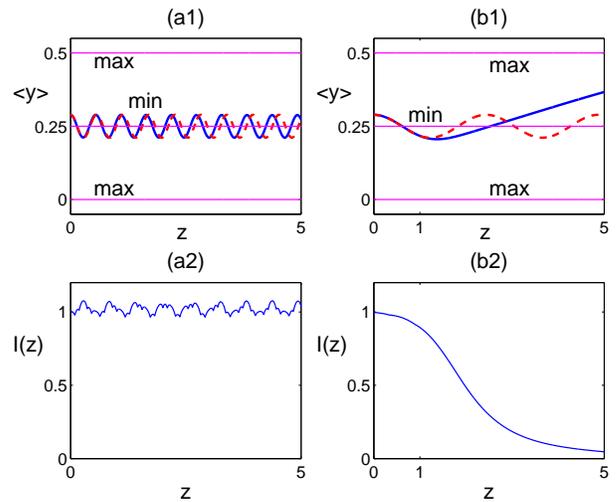
\fig{0.45}{DNS_cosmin_shift2}{DNS_cosmin_shift}{ (Color online)
 	Dynamics of solutions of Eq.~(\ref{eq:NLS}) with the periodic
	lattice~(\ref{eq:coscos}) with $V_0=5$. Initial conditions are
	position-shifted solitons [see Eq.~(\ref{eq:asym_pert})] centered at
	a lattice minimum, with $(\Delta x_0,\Delta y_0)=(0,0.04)$. 
	(a1) Center of mass in the $y$ coordinate (blue, solid line)
	and analytical prediction [Eq.~(\ref{eq:coscos_min}), red, dashes] for $\mu=-31$;
	(a2) Normalized peak intensity~(\ref{eq:peak}) for $\mu=-31$; 
	(b1) and (b2) are the same as (a1) and (a2), but for $\mu=-3$.
}

\subsection{Solitons at lattice maxima}\label{sub:coscos_max}

We now investigate solitons centered at the lattice maximum
$(x_0,y_0)=(0,0)$. Figure~\ref{fig:cos_stability_mu} shows that in general,
solitons at lattice maxima have the opposite stability
characteristics compared with those of solitons centered at lattice
minima: The slope condition is violated for narrow solitons and
satisfied for wide solitons, the power is above $P_c$~\footnote{In fact, also in this case, the soliton power approaches $g P_c$ where $g<1$~\cite{wide_solitons}, i.e., the soliton power is above $P_c$ only for solitons which are not near the band edge. },
and the perturbed-zero eigenvalues $\zev$ are always negative.
Interestingly, for the specific choice of the
lattice~(\ref{eq:coscos}), the powers and perturbed-zero eigenvalues
at lattice maxima and minima are approximately, but not exactly,
images of each other with respect to the case of a homogeneous
medium.

The negativity of the perturbed-zero eigenvalues implies that
solitons centered at a lattice maximum undergo a drift instability
(see Figure~\ref{fig:DNS_cosmax}(b)). However, if the initial
condition is subject to a power perturbation, see
Eq.~(\ref{eq:sym_pert}), then no drift occurs. In this case,
stability is determined by the slope condition. For example,
Figure~\ref{fig:stable_power} shows the dynamics of a
power-perturbed wide soliton for which the slope condition is
satisfied. When the soliton's input power is increased by $0.5\%$,
the solution undergoes small focusing-defocusing oscillations, as in
Figure~\ref{fig:DNS_cosmin}(a), i.e.,
it is stable under symmetric perturbations.
When the soliton's input power is increased by $1\%$, the
perturbation exceeds the ``basin of stability'' of the
soliton~\cite{narrow_lattice_solitons} and the soliton undergoes
collapse. These results again demonstrate that collapse and
instability are independent phenomena.

If the initial condition is asymmetric with respect to the lattice
maximum, the soliton will undergo a drift instability. In
Figure~\ref{fig:DNS_cosmax} we excite this instability with a small
upward shift, namely, Eq.~(\ref{eq:asym_pert}) with
$\shift=(0,0.02)$. Under this perturbation, the solution of
Eq.~(\ref{eq:drift}) is
\begin{equation}\label{eq:coscos_max}
\avg{x} \equiv 0, \quad \quad \avg{y} = \Delta y_0 \cdot \cosh
(\Omega_y z).
\end{equation}
with $\Omega_y \approx 3.9$. In the initial stage of the propagation
($z<0.5$) the soliton drifts toward the lattice minimum -- precisely
following the asymptotic prediction~(\ref{eq:coscos_min}), see
Figure~\ref{fig:DNS_cosmax}(a), but the soliton's amplitude is
almost constant), see Figure~\ref{fig:DNS_cosmax}(b). During the
second stage of the propagation ($0.5<z<0.99$) the soliton drifts
somewhat beyond the lattice minimum as it begins to undergo
self-focusing. In the final stage ($0.99<z<1$) the soliton undergoes
collapse (Figure~\ref{fig:DNS_cosmax}(b)). The \emph{global
dynamics} can be understood in terms of the stability conditions for
solitons centered at lattice minima and maxima as follows. The
initial soliton, which is centered at a lattice maximum, satisfies
the slope condition but violates the spectral condition. Consistent
with these traits, the soliton is focusing-stable but undergoes a
drift instability. As the soliton gets closer to the lattice
minimum, it can be viewed as a perturbed soliton centered at the
lattice minimum, for which the spectral condition is fulfilled and
the soliton power is below $P_c$ (see
Figure~\ref{fig:cos_stability_mu}(b)). Indeed, at this stage, the
drift is arrested because the beam is being attracted back towards
the lattice minimum. Moreover, the beam now is a strongly
power-perturbed soliton, since the beam power ($\approx 1.03 P_c$)
is $\approx 6\%$ above the power of the soliton at a lattice
minimum. Hence, in a similar manner to the results of
Figure~\ref{fig:DNS_cosmin}(a), the perturbation exceeds the ``basin
of stability'' and the soliton undergoes collapse.

\fig{0.5}{DNS_cosmax_stable_power}{stable_power}{
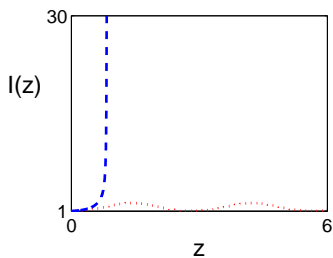
    (Color online) Same as Figure~\ref{fig:DNS_cosmin}(a)
    for a soliton at a lattice maximum with $\mu=-5$ (stable branch)
    and input power that is $0.5\%$
    (red dots) and $1\%$ (blue dashes) above the soliton power.
  }

\fig{0.5}{DNS_cosmax_shft_dx}{DNS_cosmax}{
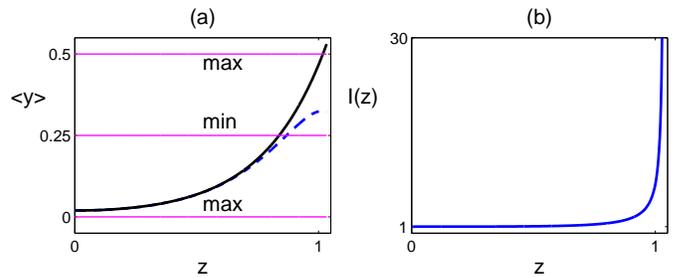
    (Color online) Dynamics of a soliton at a lattice maximum with $\mu=-5$, which is position-shifted according to~(\ref{eq:asym_pert}) with
    $\shift\approx (0,0.02)$.
    (a) Center of mass in the $y$ coordinate (blue, dashes)
    and the analytical prediction
    (Eq.~(\ref{eq:coscos_max}) with $\Omega_y \approx 3.9$, solid black). Location of lattice minimum and maxima are denoted by thin magenta and black horizontal lines, respectively. (b) Normalized peak intensity.
  }

\subsection{Solitons at a saddle point}\label{sub:coscos_saddle}

From the didactic point of view, it is interesting also to consider
solitons centered at a saddle point since they exhibit a combination
of the features of solitons at lattice minima and maxima. To show
this, we compute solitons centered at the saddle point $(x_0,y_0) =
(0.25,0)$ of the lattice~(\ref{eq:coscos}).

Figure~\ref{fig:cos_saddle}(a) shows that the zero eigenvalues
bifurcate into $\evx>0$ on the stable $x$-direction, i.e., along
direction in which the saddle is a minimum, and to $\evy<0$ on the
unstable $y$-direction, where the saddle in a maximum.

The opposite signs of the perturbed-zero eigenvalues imply a
different dynamics in each of these directions. In order to excite
only the drift instability, we solve Eq.~(\ref{eq:NLS}) with $\mu =
-12$ which belongs to the focusing-stable branch (see
Figure~\ref{fig:cos_saddle}(b2)). For this value of $\mu$, the
perturbed zero eigenvalues are $\evx \cong 1.7$ and $\evy \cong
-1.8$. By~(\ref{eq:Omega}), the theoretical prediction for the
oscillation period is $\Omega_x \cong |7i| = 7$ whereas the drift
rate is $\Omega_y \cong 7.2$. Hence, the theoretical prediction for
the dynamics of the center of mass is
$$
\avg{x} \approx 0.25 + \Delta x_0 \cdot \cos(7z), \quad \quad
\avg{y} \approx \Delta y_0 \cdot \cosh(7.2z).
$$
Indeed, a shift in the $x$ direction $\shift\approx (0.0156,0)$
leads to oscillation in the $x$-direction
(Figure~\ref{fig:DNS_saddle}(a)) while $\avg{y}$
(Figure~\ref{fig:DNS_saddle}(b)) and the amplitude
(Figure~\ref{fig:DNS_saddle}(c)) are unchanged. On the other hand, a
shift in the $y$ direction $\shift\approx (0,0.0156)$ leads to a
drift instability in the $y$-direction
(Figure~\ref{fig:DNS_saddle}(b)) but has no effect on $\avg{x}$
(Figure~\ref{fig:DNS_saddle}(a)). In both the stable and unstable
directions, the center of mass follows the analytical prediction
remarkably well.
Figure~\ref{fig:DNS_saddle}(c) also shows that once the soliton
drifts beyond the lattice minimum, the beam undergoes collapse. This
can be understood using the same reasoning used for solitons that
drift from a lattice maximum (see explanation for
Figure~\ref{fig:DNS_cosmax} in Section~\ref{sub:coscos_max}).

We also note that for the specific choice of the
lattice~(\ref{eq:coscos}), the values of the perturbed-zero
eigenvalues in the stable and unstable directions are nearly
indistinguishable from those of the perturbed-zero eigenvalues that
correspond to solitons centered at a lattice minimum and maximum,
respectively. This can be understood by rewriting the
lattice~(\ref{eq:coscos}) as
\begin{equation}
  \label{eq:coscos_saddle}
  V(x,y) = \frac{V_0}{2}\left[1 - \cos^2(2\pi (x - 0.25)) + \cos^2(2\pi y)
  \right].
\end{equation}
Thus, apart from the constant part (i.e., the first term), the
difference between the lattices is the sign before the $x$-component
of the lattice. In that sense, in the $x$ direction, the saddle
point is {\em equivalent} to a maximum point, hence, the similarity
between the eigenvalues. Another consequence of the $x-y$ symmetry
of the lattice~(\ref{eq:coscos}) is that the soliton has
approximately the critical power $P_c$ for all~$\mu$, i.e.,
$P(\mu)\approx P_c$, which is approximately the average of the
powers of solitons at maxima and minima (see
Figure~\ref{fig:cos_saddle}(b1,b2)). As noted before, this will be
no longer true if the lattice changes in the $x$ and $y$ directions
will no longer be equal.

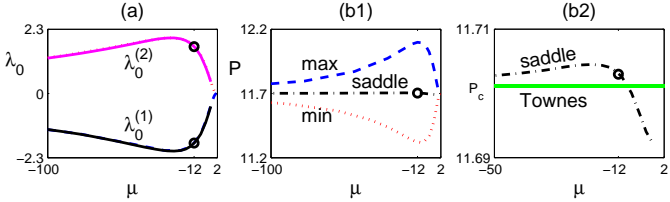
\fig{0.52}{cos_saddle}{cos_saddle}{ (Color online)
    (a) The perturbed-zero eigenvalues at the saddle point.
    One eigenvalue is shifted to positive values (magenta), and is indistinguishable
    from the eigenvalue at lattice minima (red); one eigenvalue is shifted
    to negative values (black), and is indistinguishable
    from the eigenvalue at lattice maxima (blue).
    (b1) Same data as in Figure~\ref{fig:cos_stability_mu}(a), with the addition of data for
    solitons centered at a saddle point of
    the lattice (black, dash-dots).
    (b2) same as (b1) showing only the data for
    solitons centered at a saddle point (black, dash-dots)
    and for the homogeneous medium soliton (greed line).  }

\fig{0.52}{DNS_saddle}{DNS_saddle}{
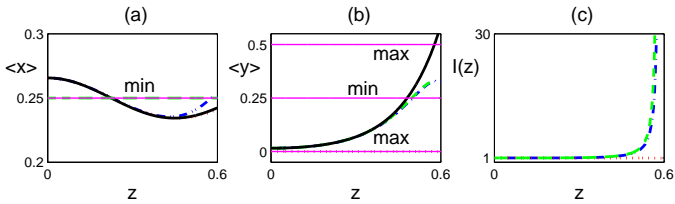
    (Color online) Dynamics of a soliton
    centered at a saddle point $(x_0,y_0) = (0.25,0)$ of the lattice~(\ref{eq:coscos})
    with $\mu=-12$ and shifts along:
    (\emph{i}) the stable $x$ direction [$\shift\approx (0.0156,0)$, red dots],
    (\emph{ii}) the unstable $y$ direction
    [$\shift\approx (0,0.0156)$, green dashes],
    (\emph{iii}) the diagonal direction
    [$\shift\approx (0.0156,0.0156)$, blue, dash-dots].
    (a) Center of mass $\avg{x}$.
    (b) Center of mass  $\avg{y}$.
    (c) Normalized peak intensity.
  }

If we apply perturbations in the stable and unstable directions
simultaneously $\shift\approx (0.0156,0.0156)$, the dynamics in each
coordinate is nearly identical to the dynamics when the
perturbation was applied just in that direction.
Thus, there is a ``decoupling'' between the (lateral dynamics in
the) $x$ and $y$ directions. Indeed, this decoupling follows
directly from Eq.~(\ref{eq:drift}).

\subsection{Solitons at a shallow-maximum}\label{sub:shallow_max}

We now consider solitons of the periodic potential
\begin{equation}
  V(x,y) = \frac{V_0}{25} \left[2 \cos(2\pi x) + 2 \cos(2\pi y) +
  1\right]^2,
\label{eq:planes}
\end{equation}
where $V_0 = 5$ and the normalization by $25$ implies that $V_0 =
\max_{x,y}V(x,y)$. Unlike the lattice~(\ref{eq:coscos}), the
lattice~(\ref{eq:planes}) also has shallow local maxima that are not
global maxima [e.g., at (0.5,0.5)].

The stability and instability dynamics of solitons centered at
global minima, maxima and saddle points of the lattice are similar
to the case of the lattice~(\ref{eq:coscos}), which was already
studied. Hence, we focus only on the stability of solitons centered
at a shallow maximum.

Since the lattice is invariant under a $90^\circ$ rotation, the
perturbed-zero eigenvalues are equal, i.e., $\evx = \evy$. However,
unlike solitons centered at a global maximum, the corresponding
perturbed-zero eigenvalues are negative only for very negative
values of $\mu$ (narrow beams) but become positive for values of
$\mu$ near the band edge $\mu_{BE}$ (wide beams), see
Figure~\ref{fig:planes_stability}(b). The reason for the positivity
of $\evx=\evy$ despite being centered at a lattice maximum is as
follows. For narrow solitons, the region where the ``bulk of the
beam'' is located is of higher values of the potential compared with
the immediate surrounding, hence, the solitons ``feel'' an effective
lattice maximum. On the other hand, for wider solitons, the ``bulk
of the beam'' is centered mostly at the shallow lattice maximum and
the surrounding lower potential regions. Hence, although the very
center of the soliton is at the shallow lattice maximum, these
solitons are effectively centered at the lattice minimum with
respect to the nearest global lattice maxima (see
also~\cite{NLS_NL_MS_1D}, Section 4.5). The transition of the
qualitative stability properties between narrow and wide solitons
described above occurs when the soliton's width is on the order of
the lattice period. As noted in
Section~\ref{sub:stability_inhom_rev}, the stability at the
transition points where $\zev = 0$ or $\dPdmu = 0$ requires a
specific study. Similarly, a comparison of
Figure~\ref{fig:planes_stability}(a) and
Figure~\ref{fig:cos_stability_mu}(a) shows that the $P(\mu)$
reflects the transition between properties which are characteristic
to solitons centered at lattice maxima and minima. Indeed, for
narrow solitons ($\mu \to - \infty$) is similar to the power of
solitons centered at a global maximum, i.e., the power is above
critical and the slope is positive. On the other hand, $P(\mu)$
curve for wide solitons ($\mu \to \mu^{(V)}_{BE}$) is similar to the
power of solitons centered at a (simple) lattice minimum, i.e., the
power is below critical and the slope is positive too.

Numerical simulations (Figure~\ref{fig:DNS_permin}) demonstrate this
transition. For a narrow soliton ($\mu = -12$), the theoretical
prediction for the dynamics of the center of mass is $\avg{x} \cong
0.5 + \Delta x_0 \cosh(4.14z)$ and $\avg{y} \cong 0.5 + \Delta y_0
\cosh(4.14z)$. Indeed, the narrow soliton drifts away from the
shallow maximum toward the nearby (global) lattice minimum
(Figure~\ref{fig:DNS_permin}(a1)) and then undergoes collapse
(Figure~\ref{fig:DNS_permin}(a2)). This dynamics is similar to that
of solitons centered near lattice maximum or a saddle of a the
lattice~(\ref{eq:coscos}), see Sections~\ref{sub:coscos_max}
and~\ref{sub:coscos_saddle}. On the other hand, for the wide soliton
($\mu = -2$), the theoretical prediction for the dynamics of the
center of mass is $\avg{x} \cong 0.5 + \Delta x_0 \cos(1.6z)$ and
$\avg{y} \cong 0.5 + \Delta y_0 \cos(1.6z)$. Indeed, this soliton
remains stable, undergoing small position oscillations around the
shallow maximum (Figure~\ref{fig:DNS_permin}(b)). This dynamics is
the same as for solitons centered at a minimum of the
lattice~(\ref{eq:coscos}), see Figure~\ref{fig:DNS_cosmin_shift}(a).
As in previous examples, the numerical results are in excellent
agreement with the analytic
prediction~(\ref{eq:drift})-(\ref{eq:Omega}).

\fig{0.52}{planes}{planes}{
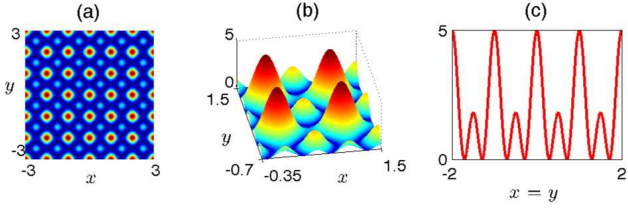
    (Color online) The shallow maximum periodic lattice given by Eq.~(\ref{eq:planes})
    with $V_0=5\,$.
    (a) Top view. (b) Side view. (c) Cross section along the line $x=y$.
  }

\fig{0.5}{planes_stability}{planes_stability}{
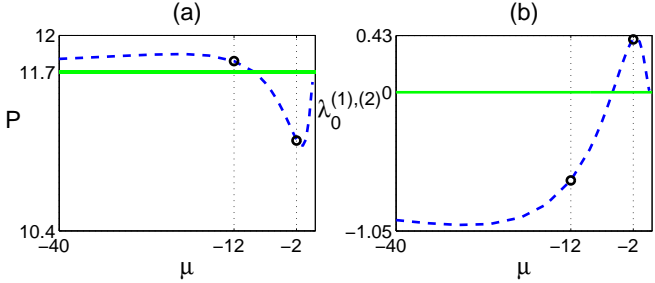
  (Color online) Same as Figure~\ref{fig:cos_stability_mu}
    for solitons centered at a
    shallow local maximum of
    the shallow-maximum periodic lattice~(\ref{eq:planes}).
  }

\fig{0.52}{DNS_permin}{DNS_permin}{
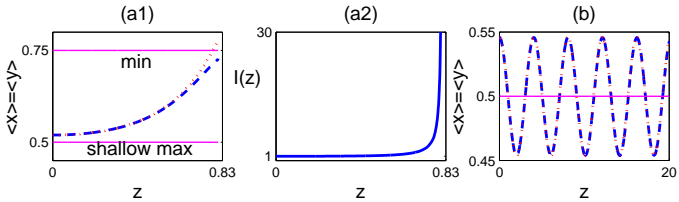
    (Color online) Dynamics of a perturbed
    soliton at shallow-maximum periodic lattice~(\ref{eq:planes})
    with a narrow soliton [(a1) and (a2) with $\mu=-12$]
    and a wide soliton [(b) with $\mu=-2$],
    and using $\shift=(0.05,0.05)$.
    (a1) Center of mass $\avg{x}=\avg{y}$ of the narrow soliton
    (blue, dashes) and the analytical prediction (red dots).
    (a2) Normalized peak intensity of the narrow soliton.
    (b) Same as (a1) for the wide soliton.
  }

\section{Periodic lattices with defects}\label{sec:defects}

Defects play a very important role in energy propagation through
inhomogeneous structures. They arise due to imperfections in natural
or fabricated media. They are also often specifically designed to
influence the propagation.

Solitons in periodic lattices with defects have drawn much attention
both experimentally and theoretically; see, for example,
~\cite{Irregular-06,Konotop_Wannier,Yang_Chen_2d_defects,GSW-JOSA-B:01}.
The complexity of the lattice details offers an opportunity to
demonstrate the relative ease of applying the stability/dynamics
criteria to predict and decipher the soliton dynamics in them. As an
example, we study lattices with a point defect. Our analysis can
also extend to different types of defects such as line defects, see
e.g.~\cite{Irregular-06}.

We consider the lattice~(\ref{eq:planes})
\begin{equation}
  V(x,y) = \frac{V_0}{25} \left| 2 \cos(2\pi x) + 2 \cos(2\pi y) +
    e^{i\theta(x,y)} \right|^2,
  \label{eq:vacancy}
\end{equation}
where the phase function $\theta(x,y)$ is given by
\begin{equation}
  \label{eq:phase_vacancy}
  \theta(x,y) = \tan^{-1}\left(\frac{y-y_0}{x}\right) -
  \tan^{-1}\left(\frac{y + y_0}{x}\right),
\end{equation}
see Figure~\ref{fig:vacancy} and also~\cite{Irregular-06}. Compared
with the shallow-maximum periodic lattice~(\ref{eq:planes}), here
the constant (DC) component (the third term in the lattice) attains
a phase distortion which creates an (effective) vacancy defect at
$(0,0)$, which is a shallow-maximum. Further, far away from the
origin, the potential~(\ref{eq:vacancy}) is locally similar to the
shallow-maximum periodic lattice~(\ref{eq:planes}). This is a
generic example of a {\em point} defect, as opposed to a {\em line}
defect~\cite{NP_vacancy}. In what follows, we consider solitons
centered at the vacancy defect $(x_0,y_0) = (0,0)$.

The stability properties of solitons in the shallow-maximum
periodic~(\ref{eq:planes}) and vacancy-defect~(\ref{eq:vacancy})
lattices are strikingly similar, as can be seen from
Figs.~\ref{fig:planes_stability} and~\ref{fig:vac_stability}. In
both cases, there is a marked transition between narrow and wide
solitons and this transition occurs when the soliton width is of the
order of the lattice period. Indeed, numerical simulations show that
the dynamics of perturbed solitons is qualitatively similar in both
cases -- compare Figures~\ref{fig:DNS_permin}
and~\ref{fig:DNS_vacancy}. We do note that unlike the
shallow-maximum periodic lattice, the perturbed-zero eigenvalues of
the vacancy lattice bifurcate into different, though similar,
values. The reason for this is the phase
function~(\ref{eq:phase_vacancy}) is not invariant by $90^\circ$
rotations.

Inspecting the lattice surfaces (Figures~\ref{fig:planes}
and~\ref{fig:vacancy}), it is clearly seen that the reason for the
similarity between the shallow-maximum periodic and vacancy lattices
is that the vacant site is essentially a shallow local maximum
itself -- and only a bit shallower than those of the shallow-maximum
periodic lattice (see Figure~\ref{fig:vacancy}).

In Figure~\ref{fig:vacancy_drift_collapse} we give a detailed
graphical illustration of a typical instability dynamics due to a
violation of the spectral condition.
Figure~\ref{fig:vacancy_drift_collapse}(a)-(c) show contours of the
soliton profiles superposed on the contour plot of the lattice. It
can be seen that as a result of the initial position shift, the
soliton drifts towards the lattice minimum and that it self-focuses
at the same time. Figure~\ref{fig:vacancy_drift_collapse}(d) shows
the trajectory of the beam across the lattice. In addition,
Figure~\ref{fig:vacancy_drift_collapse}(e) shows the center of mass
dynamics as a function of the intensity $I(z)$. This shows that
initially, the perturbed soliton undergoes a drift instability with
little self-focusing, but that once the collapse accelerates, it is
so fast so that the drift dynamics becomes negligible.

\fig{0.52}{vacancy}{vacancy}{
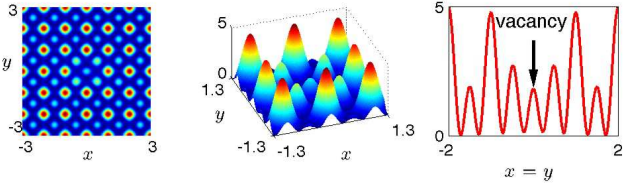
  (Color online) Same as Figure~\ref{fig:planes} for solitons centered at the
  ``vacancy'' of the lattice~(\ref{eq:vacancy}).
}

\fig{0.5}{vac_stability}{vac_stability}{
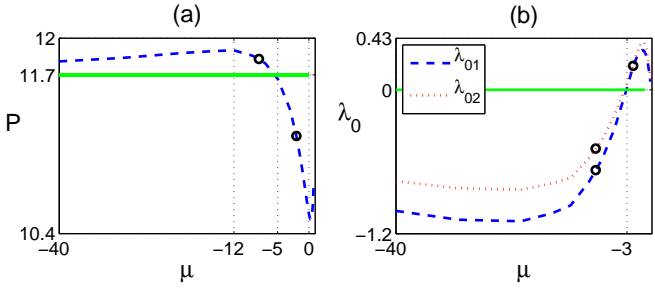
  (Color online) Same as Figure~\ref{fig:cos_stability_mu} for solitons at the
    vacancy of the lattice~(\ref{eq:vacancy}).
    (b) The perturbed-zero eigenvalues $\both$ are slightly different
    from each other.
    The circles (black) correspond to the values used
    in Figure~\ref{fig:DNS_vacancy}.
}

\fig{0.52}{DNS_vacancy}{DNS_vacancy}{
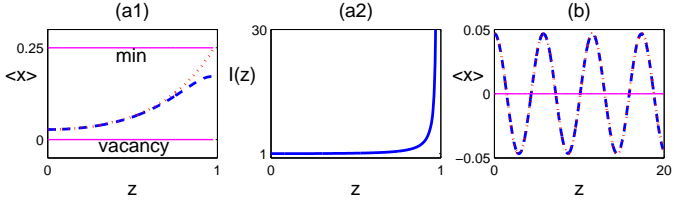
    (Color online) Same as Figure~\ref{fig:DNS_permin} but
    for the vacancy lattice~(\ref{eq:vacancy}).
    Here $\Omega_x\approx 3$ in (a2) and
    $\Omega_x\approx 1.09i$ in (b).
    In both cases the $\avg{y}$ dynamics (not shown)
    is similar (but not identical) to the $\avg{x}$ dynamics.
  }

\fig{0.52}{vacancy_drift_collapse}{vacancy_drift_collapse}{
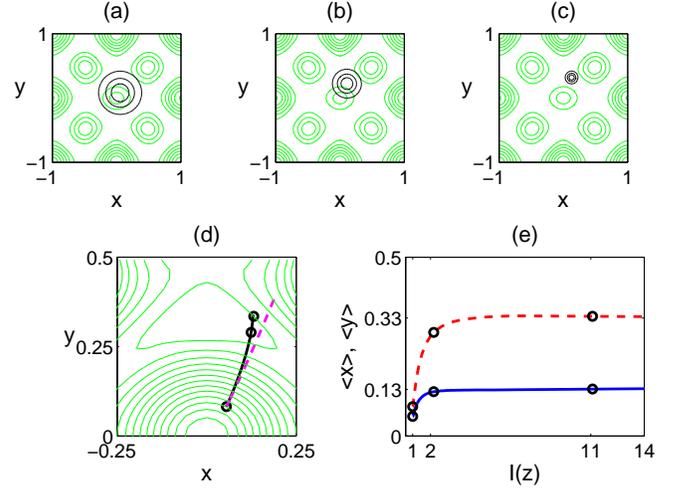
(Color online)
(a)--(c): contours of the intensity~$|u(x,y,z)|^2$ (blue)
superimposed on the vacancy lattice (green) with initial conditions
corresponding to the mode with $\mu = -8$ that is initially shifted
in the $(x,y)$ plane to $\shift = (0.05,0.1)$, i.e., at an angle of
$63^o$ to the $y$ axis.
(a) $z=0,~I \approx 1$,
(b) $z=0.51,~I
\approx 2.18$,
(c) $z=0.63,~I \approx 11.1$.
(d) Center of mass dynamics (black curve) and the
analytical prediction (magenta, dashes)
superimposed on the contours of the potential (green).
(e) $\avg{x}$ (blue, solid) and $\avg{y}$ (red, dashes)
as functions of $I(z)$.
Circles (black) correspond to the z-slices shown in (a)--(c).
}

\section{Quasicrystal lattices}\label{sec:quasicrystal}
Next, we investigate solitons in quasicrystal lattices. Such
lattices appear naturally in certain
molecules~\cite{Shechtman-84,Marder-01}, have been investigated in
optics~\cite{Bratfalean-05,Lifshitz-05,Man-05,Freedman-06,Irregular-06}
and in BEC~\cite{Palencia-05}, and can be formed optically by the
far-field diffraction pattern of a mask with point-apertures that
are located on the $N$ vertices of a regular polygon, or
equivalently, by the sum of $N$ plane waves
(cf.~\cite{Berry-74,Irregular-06}) with wavevectors $(k_x,k_y)$
whose directions are equally distributed over the unit circle. The corresponding
potential is given by
\begin{equation}
  \label{eq:QC-N}
     V(x,y) = \frac{V_0}{N^2}\left|\sum_{n=0}^{N-1}e^{i(k^{(n)}_x x + k^{(n)}_y y)}\right|^2~,
\end{equation}
where $(k^{(n)}_x,k^{(n)}_y)=\left(K\cos(2\pi n/N),K\sin(2\pi
n/N)\right)$~\footnote{We note that Eq.~(\ref{eq:QC-N}) can also
describe the lattices~(\ref{eq:planes}) and~(\ref{eq:vacancy}) for
$N=4$ and an additional $k=0$ phase modulated plane wave. }. The
normalization by $N^2$ implies that $V_0 = \max_{x,y}V(x,y)$. The
potential~(\ref{eq:QC-N}) with $N=2,3,4,6$ yields periodic lattices.
All other values of $N$ correspond to quasicrystals, which have a
local symmetry around the origin and long-range order, but, unlike
periodic crystals, are not invariant under spatial
translation~\cite{Senechal-95}.

We first consider the case $N=5$ (a 5-fold symmetric ``Penrose''
quasicrystal) for solitons centered at the lattice maximum
$(x_0,y_0) = (0,0)$, see Figure~\ref{fig:penrose5}. Since the
soliton profile and stability are affected mostly by the lattice
landscape near its center, we can expect the stability properties of
the Penrose lattice soliton at $(0,0)$ to be qualitatively the same
as for a soliton at a lattice maximum of a periodic lattice. Indeed,
Figure~\ref{fig:penrose_stability} reveals the typical stability
properties of solitons centered at a lattice maximum: An
focusing-unstable branch for narrow solitons, an focusing-stable
branch for wider solitons and negative perturbed zero-eigenvalues
(compare e.g. with Figure~\ref{fig:DNS_cosmax}). Therefore, the
Penrose soliton will drift from the lattice maximum under asymmetric
perturbations and if the soliton is sufficiently narrow, it can also
undergo collapse.

Figure~\ref{fig:penrose_stability} presents also the data for a
perfectly periodic lattice ($N=4$) and for a higher-order
quasicrystal ($N=11$). One can see that the stability properties in
these lattices is qualitatively similar to the $N=5$ case. The only
marked difference as $N$ increases is that the soliton's power
becomes larger for a given $\mu$.

These results show that in contrast to the significant effect of the
quasi-periodicity on the dynamics of linear waves (compared with the
effect of perfect periodicity~\cite{Freedman-06}),
the effect of quasi-periodicity on the dynamics of solitons is small.

\fig{0.52}{penrose5new}{penrose5}{
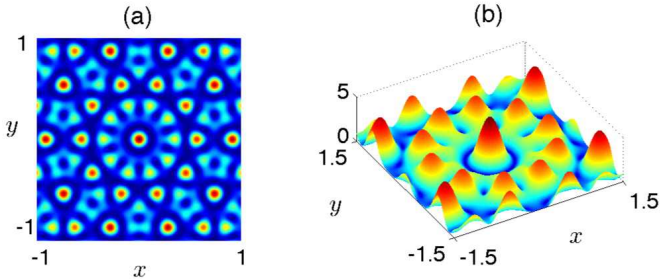
    (Color online) Same as Figure~\ref{fig:planes}(a)+(b) for the
    Penrose quasicrystal
    lattice given by Eq.~(\ref{eq:QC-N}) with
    $N=5$ and $V_0=5$.
    }

\fig{0.5}{penrose_stability}{penrose_stability}{
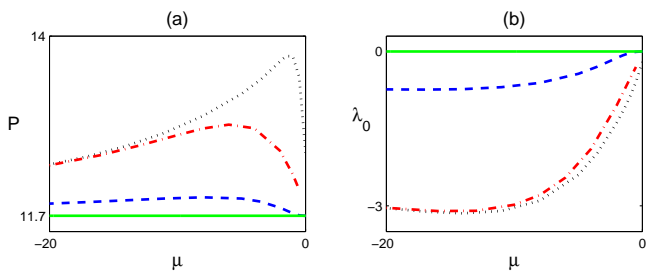
  (Color online) Same as Figure~\ref{fig:cos_stability_mu} for solitons at
  the maxima of the lattices~(\ref{eq:QC-N}) with
  $N=4$ (periodic lattice, dashed blue line), $N=5$ (Penrose quasicrystal lattice, dash-dotted red line), $N=11$ (higher-order quasicrystal
  lattice, dotted black line), and the homogeneous NLS soliton (solid green line).
}

\section{Single waveguide potentials}

So far we studied periodic, periodic potentials with defects and
quasiperiodic potentials. However, our theory can be applied to
other types of potentials. Indeed, let us consider localized
potentials, such as single or multiple waveguide potentials, for
which the potential decays to zero at infinity. For such potentials,
there are two limits of interest. The first limit is of solitons
which are much wider than the width of the potential. In this case,
the potential can be approximated as a point defect in an
homogeneous medium. Then, the dynamics is governed by
\begin{equation}
\label{eq:NLS_delta}
   i A_z(\vec{x},z) + \Delta A + |A|^{2\sigma}A - \gamma\delta(\vec{x})A =
   0,
\end{equation}
where $\gamma$ is a real constant. In~\cite{delta_pot_complete}, the
qualitative and quantitative stability approaches were applied to
Eq.~(\ref{eq:NLS_delta}) in one transverse dimension.

The second limit is of solitons which are much narrower than the
width of the potential. In this case, only the local variation of
the potential affects the soliton profile and stability. Hence, the
potential can be expanded as
$$
V(x) = V(0) + \frac{1}{2}V''(0)x^2 + \cdots.
$$
The qualitative and quantitative stability approaches were applied
to this case in~\cite{narrow_lattice_solitons}.

In~\cite{delta_pot_complete,narrow_lattice_solitons}, the profiles,
power slope and perturbed-zero eigenvalues were computed
analytically (exactly or asymptotically). It was proved that the
perturbed-zero eigenvalues are negative for solitons centered at
lattice maxima (repulsive potential) and are positive for solitons
centered at lattice minima (attractive potential). Hence, in the
latter case, stability is determined by the slope condition. In
those two studies, detailed numerical simulations confirmed the
validity of the qualitative and quantitative approaches. Hence, we
do not present a systematic stability study for localized
potentials.

\section{Final remarks}
In this paper, we presented a unified approach for analyzing the
stability and instability dynamics of
positive bright solitons. This approach consists of a
{\em qualitative} characterization of the type of instability, and a
{\em quantitative} estimation of the instability growth rate and the
strength of stability. This approach was summarized by several rules
(Section~\ref{sec:rules}) and applied to a variety of numerical
examples (Sections~\ref{sec:periodic}-\ref{sec:quasicrystal}), thus
revealing the similarity between a variety of physical
configurations which, {\em a priori}, look very different from each
other. In that sense, our approach differs from most previous
studies which considered a specific physical configuration.

One aspect which was emphasized in the numerical examples is the
excellent agreement between direct numerical simulations of the NLS
and the reduced equations for the center of mass (lateral) dynamics,
Eqs.~(\ref{eq:drift})-(\ref{eq:Omega}). Different reduced equations
for the lateral dynamics were previously derived under the
assumption that the beam remains close to the initial soliton
profile (see e.g.~\cite{kartashov_1d}) or by allowing the soliton
parameters to evolve with propagation distance (see
e.g.,~\cite{Agrawal-Kivshar-book} and references therein). These
approaches, as well as ours, are valid only as long as the beam
profile remains close to a soliton profile. However, unlike previous
approaches, Eqs.~(\ref{eq:drift})-(\ref{eq:Omega}) incorporate
linear stability (spectral) information into the center of mass
dynamics. Thus our approach shows that the beam profile evolves as a
soliton perturbed by the eigenfunction $f_{0,j}^{(V)}$. The validity
of this perturbation analysis is evident from the excellent
comparison between the reduced
Eqs.~(\ref{eq:drift})-(\ref{eq:Omega}) and numerical simulations for
a variety of lattice types. To the best of our knowledge, such an
agreement was not achieved with the previous approaches.

The numerical examples in this paper were for two-dimensional Kerr
media with various linear lattices. Together with our previous
studies which were done for narrow solitons in any
dimension~\cite{narrow_lattice_solitons}, a linear delta-function
potential~\cite{delta_pot_complete} and for nonlinear
lattices~\cite{NLS_NL_MS_1D,NLS_NL_MS_2D}, there is a strong
numerical evidence that our qualitative and quantitative approaches
apply to positive solitons in any dimension, any type of
nonlinearity of type $F(|A|^2)$ (e.g., saturable) as well as for other lattice
configurations, e.g., ``surface'' or ``corner''
solitons~\cite{discrete_surface_solitons}.

Theorem~\ref{th:stability} as well as the qualitative and
quantitative approaches apply also for the $d$-dimensional discrete
NLS. This equation is obtained from Eq.~(\ref{eq:NLS}) by replacing
$\Delta$ by the difference Laplacian operator on a discrete lattice
and $V$ by a potential defined at discrete lattice sites. This model
was extensively studied, mostly for periodic lattices, see e.g., for
1D and 2D discrete NLS equation with cubic nonlinearity (see
e.g.,~\cite{discrete_solitons,eisen-prl1998,efrem-pre2002pr,Christodoulides-03,Fleischer-03,Sukhorukov-03}),
saturable nonlinearity (see e.g.,~\cite{discrete_NLS_saturable}),
cubic-quintic nonlinearity (see e.g.,~\cite{Malomed-discrete-cq}).
General results on existence and stability of solitons in $d-$
dimensions with power nonlinearities appear in
\cite{Weinstein-99,Weinstein-discrete}. Indeed, for the discrete
NLS, the operator $L_+$ does not generically have a zero eigenvalue
due to absence of continuous translation symmetry, and the
continuous spectrum is a bounded interval, starting at the soliton
frequency, $-\mu$~\cite{Weinstein-discrete}. However, these changes
in the spectrum do not affect the stability theory, the possible
types of instabilities and the analysis of their strength.

As noted, our analysis shows that for positive bright solitons, only
two types of instabilities are possible - focusing instability or
drift instability. Other types of instabilities may appear, but only
for non-positive solitons (e.g., gap solitons or vortex solitons). A
formulation of a qualitative and quantitative theories for such
solitons requires further study.

\section*{Acknowledgments} We acknowledge useful
discussions with M.J. Ablowitz. The research of Y. Sivan and G.
Fibich was partially supported by BSF grant no. 2006-262. M.I.
Weinstein was supported, in part, by US-NSF Grants DMS-04-12305 and
DMS-07-07850.

\bibliographystyle{apsrev}

\end{document}